\documentclass[12pt, twoside]{article}
\usepackage{a4wide,amssymb,cite}
\usepackage{epsfig}

\newcommand{\be}{\begin{equation}}
\newcommand{\ee}{\end{equation}}
\newcommand{\bea}{\begin{eqnarray}}
\newcommand{\eea}{\end{eqnarray}}

\def\de{\partial}
\def\a{\alpha}
\def\b{\beta}
\def\g{\gamma}
\def\G{\Gamma}
\def\d{\delta}
\def\D{\Delta}

\def\la{\lambda}
\def\La{\Lambda}

\def\m{\mu}
\def\n{\nu}
\def\r{\rho}
\def\o{\omega}
\def\s{\sigma}

\def\vf{\varphi}
\def\ep{\epsilon}
\def\vep{\varepsilon}

\def\th{\theta}
\def\Th{\Theta}
\def\de{\partial}

\newcommand{\eg}{{\it e.g.,}\ }\newcommand{\ie}{{\it i.e.,}\ }
\newcommand{\reef}[1]{(\ref{#1})}

\begin{document}

\begin{titlepage}

\rightline{May 2007}

\begin{centering}
\vspace{1cm}
{\Large {\bf Gravitino in six-dimensional warped supergravity}}\\

\vspace{1.5cm}

 {\bf Hyun Min Lee}$^{a,*}$ and {\bf Antonios~Papazoglou} $^{b,c,**}$ \\
\vspace{.2in}

$^{a}$ Department of Physics, Carnegie Mellon University, \\
5000 Forbes Avenue, Pittsburgh, PA 15213, USA.\\
\vspace{3mm}
$^{b}$ APC\footnote{UMR 7164(CNRS, Universit\'e Paris 7, CEA, Obervatoire de Paris)}, 10 rue Alice Domon et L\'eonie Duquet, \\
75205 Paris Cedex 13, France. \\
\vspace{3mm}
$^{c}$ GReCO/IAP\footnote{UMR 7095(CNRS, Universit\'e Paris 6)}, 98 bis Boulevard Arago, \\
75014 Paris, France.

\end{centering}
\vspace{2cm}

\begin{abstract}
We consider the gravitino spectrum for the general warped solution
in a specific six-dimensional gauged supergravity. We find that
although the brane tensions introduced at the conical
singularities break the bulk supersymmetry explicitly, massless
modes of the gravitino can exist with a nontrivial wave function
profile, due to a nonzero $U(1)_R$ gauge flux. We also compute the
wave function and the mass spectrum of Kaluza-Klein massive modes
of the gravitino explicitly. We show that the introduction of a
 gravitino mass term on a  regularized brane can give a suppressed
effective gravitino mass compared to the compactification scale,
due to the delocalization of the wave function of the zero-mode
gravitino.

\end{abstract}

\vskip 1cm


\vspace{2cm}
\begin{flushleft}

$^{*}~$ e-mail address: hmlee@andrew.cmu.edu \\
$ ^{**}$ e-mail address: papazogl@iap.fr

\end{flushleft}
\end{titlepage}

\section{Introduction}

%motivation
There has been a lot of interest in brane world models in higher
dimensions with the hope to solve the particle physics problems
and give a hint for physics beyond the Standard Model (SM) in a
different context. Particularly, in order to ameliorate the
hierarchy problem of the Higgs mass, models with extra dimensions
compactified on a large flat \cite{add} or small warped \cite{rs}
space were suggested as an alternative to the weak-scale
supersymmetry (SUSY). Furthermore, regarding the cosmological
constant problem which has been one of the most notorious problems
as dictated by a no-go theorem \cite{weinberg}, the self-tuning
mechanism in higher dimensions \cite{selftune} was suggested. This
 may give a better understanding of the cosmological constant
problem, although one has only the SM quantum corrections confined
on a brane under control. In particular, brane world models in six
dimensions have drawn much attention because the brane tension
generates a nonzero deficit angle in extra dimensions without
curving the 4D spacetime \cite{deficit}. This feature has been
first pursued in   the framework of  spontaneous compactification
due to gauge fluxes in 6D Einstein-Maxwell theory \cite{6dself},
but ended up with a fine-tuning condition for the brane tension
due to flux quantization or conservation \cite{finetune}.
Furthermore, we still need some symmetry to ensure that a bulk
tuning condition is stable against the quantum corrections.

%6D supergravity
The Salam-Sezgin (SS) supergravity \cite{SS} has drawn a renewed
interest due to the possibility of attacking both brane and bulk
fine-tuning problems encountered in the non-supersymmetric models.
In this model, Salam and Sezgin obtained a spontaneous
compactification on a sphere with $U(1)_R$ flux to get the 4D
Minkowski spacetime and showed that 4D ${\cal N}=1$ SUSY survives
or there is a massless chiral gravitino in four dimensions. The
most general warped non-singular (no worse than conical) solutions
with 4D maximal symmetry have been recently found to be a warped
product of the 4D Minkowski space and a two dimensional compact
manifold (the ``football-shaped'' space \cite{burgess1}, the general warped solution with axial symmetry \cite{gibbons} or the general warped solution without axial symmetry \cite{lee}).
Nonetheless, there is still a
fine-tuning between brane tensions due to the flux quantization.
It has been shown, on the other hand,  that there are warped
singular solutions with 4D curved spacetime \cite{singularsol}.
The stability analysis of the warped background has been done for
scalar perturbations \cite{scalarpert} and bulk gauge fields and
fermions \cite{salvio}.

%susy breaking and gravitino mass

The warping requires  the existence of conical singularities on
which codimension-two branes are located, so ${\cal N}=1$ SUSY is
broken explicitly by nonzero brane tensions \footnote{Such explicit SUSY breaking localized terms are a usual addition in supergravity models, \eg $\overline{D3}$ branes in throat geometries \cite{KKLT}.}.  Nevertheless, it has
been suggested that the SUSY breakdown at the branes might be
suppressed not to give a large quantum correction in the bulk
\cite{bulkcorr}. The Casimir effect has been discussed for flux compactifications
in non-SUSY and SUSY models \cite{mina}.
We will, thus, discuss the model at the classical level, not taking into account the transmission of the SUSY breaking from the brane sources to the bulk. In this paper, we study the gravitino equation
for the general warped solution in an anomaly free model with the
gauge group $E_6\times E_7\times U(1)_R$ in six-dimensional gauged
supergravity \cite{RSS}\footnote{For recent studies of
six-dimensional supergravities, see Ref.~\cite{moresugra}.}. The
difference from the SS model is that on top of the $U(1)_R$ flux,
we can also turn on the abelian flux of the non-abelian gauge
fields, for instance, $E_6$, but the form of the general warped
solution is maintained. In this case, both SUSY and the GUT group
can be  broken upon the compactification so the model
could provide an alternative to orbifold models. It has been shown
that $E_6$ is broken down to $SO(10)$ and the adjoint fermions of
$E_6$ survive as two chiral ${\bf 16}$'s of $SO(10)$ \cite{RSS}.

By solving the gravitino  equation, we find that there exist
massless modes of the gravitino for a nonzero $U(1)_R$ flux, even
with nonzero brane tensions and/or abelian flux of $E_6$. We show
that the wave function of the massless modes of the gravitino have
a nontrivial profile in the extra dimensions. Moreover, we also
obtain the wave function and the mass spectrum for the massive
modes of the gravitino. Although we can show from the bulk  SUSY
transformation that SUSY is broken for the warped solution, the
appearance of massless modes of the gravitino tells us that the
gravitino mass must be determined by interaction terms beyond the
bulk action and the brane tensions.

However, we cannot consider brane matter terms on the
codimension-2 branes without introducing singularities
 \cite{ClineGiova}. It is imperative that the brane is regularized
 by either acquiring some thickness  \cite{thickb}, or by the
 reduction of its codimension \cite{tps,ppz}. Considering the
 latter method (codimension reduction) and a
  brane-induced gravitino mass term on the regularized brane, we found
  that the gravitino mass can be
suppressed compared to the compactification scale due to the
delocalization of the zero-mode gravitino.

The paper is organized as follows. We first review the warped
solution in 6D gauged supergravity. Then, we consider the SUSY
transformation and the gravitino action in the warped background.
Further, we present the wave functions of the massless modes of
the gravitino and the mass spectrum of KK massive modes of the
gravitino. Next, we introduce brane-induced gravitino mass terms
and estimate the effective gravitino mass related to the value of
 the wave function of the zero-mode gravitino on the brane.
Finally, the conclusions are drawn.

\section{The model}

We consider an anomaly-free model\cite{RSS} in 6D gauged
supergravity where the bulk gauge group is $E_6\times E_7\times
U(1)_R$ with gauge couplings $g_6$, $g_7$ and $g_1$. By setting
the Kalb-Ramond field and the hyperscalars to zero, the bosonic
part of the bulk action\footnote{For comparison with the
Salam-Sezgin \cite{SS} notation, we note that $g_{SS}=2g_1,
A^{SS}_M=A_M/2$, $\sigma=\phi/2$, and ${\cal L}_{SS}={\cal L}/4$.
We set the 6D fundamental scale to $\kappa=1$.}
 is given by \bea e^{-1}{\cal
L}_b=R-\frac{1}{4}(\partial_M\phi)^2-8g^2_1 e^{-\frac{1}{2}\phi}
-\frac{1}{4}e^{\frac{1}{2}\phi}(F^2_{MN}+F^{I2}_{6MN}+F^{I'2}_{7MN}).
\eea When all non-abelian gauge fields vanish, the above action
becomes the one of the Salam-Sezgin model\cite{SS}. The case with
only the $U(1) \subset E_6$ being nonzero was considered for a GUT
breaking in \cite{RSS}. In the present paper, we take the more
general case where both $U(1)_R$ and $U(1)$ are nonzero. In this
case, it has been known that supersymmetry is broken completely
and $E_6$ is reduced to $SO(10)$\cite{RSS}.

Assuming axial symmetry in the internal space, the general warped
solution with $U(1)_R$, $U(1)$ fluxes can be found to give rise to four-dimensional Minkowski vacua (\ie with zero effective cosmological constant) with the following form \cite{gibbons}
\bea
ds^2&=&W^2(r)
\eta_{\mu\nu}dx^\mu dx^\nu+R^2(r)\bigg(dr^2
+\lambda^2 \Th^2(r)d\theta^2\bigg), \\
F_{1r\theta}&=&\lambda q  \frac{\Th R^2}{W^6} \cos\alpha  , \ \
T_IF^I_{6r\theta}=T_0\lambda q   \frac{\Th R^2}{W^6} \sin\alpha ,  \label{flux}\\
\phi&=&4\ln W , \eea with \bea R={W \over f_0}, \ \  \ \Th={r \over W^4},   \\
W^4=\frac{f_1}{f_0}, \ \ f_0=1+\frac{r^2}{r^2_0}, \ \ \
f_1=1+\frac{r^2}{r^2_1}, \eea where $q$ is a  constant denoting
the magnetic flux, $T_0$ is the $U(1)$ generator of $E_6$ and the
two radii are given by \be r^2_0=\frac{1}{2g^2_1}, \ \
r^2_1=\frac{8}{q^2}. \ee The angle $\a$ denotes the distribution
of the fluxes in the $U(1)_R$ and the $U(1)$ directions. From
eq.~(\ref{flux}), we get the nonzero component of gauge fields in
the patch including $r=0$ as \be g_1 A_{1\theta}=-n_1
\bigg(\frac{1}{f_1}-1\bigg), \ \ g_6
 T_IA^I_{6\theta}=-T_0 n_6 \bigg(\frac{1}{f_1}-1\bigg).
\ee where we have taken into account the quantization conditions
of the gauge fluxes  \be \frac{4\lambda g_1\cos\alpha}{q}=n_1, \ \
\frac{4\lambda g_6\sin\alpha}{q}=n_6, \ \ n_1,n_6={\rm integer}.
\ee In this general solution, the metric has two conical
singularities, one at $r=0$ and the other at $r=\infty$, with
deficit angles $\delta_s$ (supported by brane tensions
$V_s=2\delta_s$) given by
\bea
\frac{\delta_0}{2\pi}&=&1-\lambda, \\
\frac{\delta_\infty}{2\pi}&=&1-\lambda\frac{r^2_1}{r^2_0}=1-\frac{1}{\lambda}\Big(n^2_1+\frac{g^2_1}{g^2_6}n^2_6\Big).
\eea
These brane terms, as it will be discussed later, are explicitly non-supersymmetric. Let us note also, that the gauge field Bianchi identities are satisfied as long as there is no coupling of the brane to the bulk gauge field \cite{burgess1}.

 For $r_0=r_1$, {\it i.e.} $q=4g_1$, we get the unwarped  solution
with football shaped extra dimensions. In this case, we get
$\lambda^2=n^2_1+ n_6^2 g^2_1/g^2_6$ and if both $n_1$ and $n_6$
are nonzero, the angle deficit has to be negative (and so has to
be the brane tension). If additionally $\lambda=1$, the unwarped
solution is possible with no branes present. The latter happens
only for the following two cases: $(n_1,n_6)=(1,0)$ or
$(n_1,n_6)=(0,g_6/g_1)$. So, in this case, only either of $U(1)$
fluxes can be nonzero.

Finally, by defining \be d\rho=R dr, \ \ a=\lambda R \Th, \ee the
metric can be  expressed in a Gaussian normal coordinate system as
\be ds^2=W^2\eta_{\mu\nu}dx^\mu dx^\nu +d\rho^2+a^2d\theta^2. \ee

\section{Explicit supersymmetry breaking and gravitino dynamics in the warped background}

In this section we will discuss the gravitino spectrum in the
general warped background and the existence of gravitino zero
modes even in some cases where supersymmetry is broken by the
presence of the brane tension terms. In order to do this analysis,
we need the spinor part of the action and in particular the part
that is quadratic in fermionic terms. This is given by
\cite{RSS}\footnote{In comparison with the  SS notation \cite{SS},
all fermions are rescaled as $\psi_{SS}=\psi/2$.}
 \bea e^{-1}{\cal
L}_f&=&{\bar\psi}_M\Gamma^{MNP}{\cal D}_N\psi_P+
{\bar\chi}\Gamma^M{\cal D}_M\chi
+{\bar\lambda}\Gamma^M{\cal D}_M\lambda+{\bar\lambda}_6\Gamma^M{\cal D}_M\lambda_6\nonumber\\
&&+\frac{1}{4}(\partial_M\phi)({\bar\psi}_N\Gamma^M\Gamma^N\chi +
{\rm
h.c.})+\sqrt{2}g_1e^{-\frac{1}{4}\phi}(i{\bar\psi}_M\Gamma^M\lambda_1
-i{\bar\chi}\lambda_1+{\rm h.c.})  \nonumber \\
&&-\frac{1}{4\sqrt{2}}e^{\frac{1}{4}\phi}\bigg\{F_{1MN}({\bar\psi}_Q\Gamma^{MN}\Gamma^Q
\lambda_1+{\bar\chi}\Gamma^{MN}\lambda_1)  \nonumber \\
&&~~~~~~~~~~~~~~~~~~~~~~~~~
+F^I_{6MN}({\bar\psi}_Q\Gamma^{MN}\Gamma^Q
\lambda^I_6+{\bar\chi}\Gamma^{MN}\lambda^I_6) +{\rm h.c.}
\bigg\},~~~~~~ \label{faction} \eea where the covariant derivative
for any fermion $\psi$ is defined as \be {\cal
D}_M\psi=(\partial_M+\frac{1}{4}\omega_{MAB}\Gamma^{AB}-ig_1A_M)\psi.
\ee

The above spinors are chiral with handednesses \be \G^7 \psi_M = +
\psi_M, \ \ \  \G^7 \chi = - \chi, \ \ \  \G^7 \la_1 = + \la_1, \
\ \  \G^7\la_6 = + \la_6. \ee Taking into account that $\G^7=\s^3
\otimes {\bf 1}$ (see Appendix A), the 6D (8-component) spinors
can be decomposed to 6D Weyl (4-component) spinors as \be \psi_M=
(\tilde{\psi}_M,0)^T, \ \ \  \chi =(0,\tilde{\chi})^T, \ \ \ \la_1
= (\tilde{\la}_1,0)^T, \ \ \ \la_6 =(\tilde{\la}_6,0)^T. \ee

\subsection{Supersymmetry transformations}

Before presenting the gravitino spectrum, let us discuss the
supersymmetry of the vacua in question. For the general background
with fluxes, the nontrivial bulk supersymmetry transformations  of
fermions are the ones for dilatino $\chi$, gravitino $\psi_M$,
$U(1)_R$ gaugino $\lambda_1$ and $E_6$ gauginos
$\lambda^a_6$\cite{RSS}:
\bea
\delta\chi&=&-\frac{1}{4}(\partial_m\phi)\Gamma^m\varepsilon, \label{chivar} \\
\delta\lambda_1&=&\frac{1}{4\sqrt{2}}
e^{\frac{1}{4}\phi}F_{1mn}\Gamma^{mn}\varepsilon
-i\sqrt{2}g_1 e^{-\frac{1}{4}\phi}\varepsilon, \label{la1var} \\
\delta(T_I\lambda^I_6)&=& \frac{1}{4\sqrt{2}}e^{\frac{1}{4}\phi}
(T_IF^I_{6mn})\Gamma^{mn}\varepsilon, \label{la6var} \\
\delta\psi_M&=&{\cal D}_M\varepsilon, \label{psivar}
\eea where
the spinor parameter $\vep$ is chiral, with $\G^7 \vep  =  +
\vep$, so that $\vep= (\tilde{\vep},0)^T$.

Using the solution that we presented in the previous section, the above transformations give
\bea
\d \chi &=& -{W' \over W}\left[ \cos\th ~\s^1 \otimes \g^5 + \sin\th  ~\s^2 \otimes {\bf 1}   \right] \vep, \label{chiexp}\\
\d \la_1 &=& { i \over \sqrt{2} W}2 g_1 \s^3 \otimes \left[{q \over 4 g_1} {1 \over W^4} \cos\a ~\g^5 - 1 \right] \vep, \\
\delta(T_I\lambda^I_6)&=&T_0~ i \sqrt{2}~ {q \over 4} {1 \over W^4}\sin\a~ \s^3 \otimes \g^5 \vep, \\
\d \psi_\m &=& \left[ \de_\m +{1 \over 2}W' \s^3 \otimes \g_\m (\cos\th~\g^5+i \sin\th) \right] \vep, \\
\d \psi_\r &=& \de_\r \vep, \\
\d \psi_\th &=& \left[\de_\th  +{i \over 2}\left(1+{\la \over W^4}(1-{2 \over f_0})+3\la r{W' \over W^5}\right) \s^3 \otimes \g^5 + i \la{4 g_1 \over q}\cos\a \left({1 \over f_1}-1\right) \right] \vep. \label{psithexp}~~~~~~
\eea

In particular, for the sphere solution with only $U(1)_R$ flux
turned on, {\it i.e.} the Salam-Sezgin vacuum with $\la=1$,
$(n_1,n_6)=(1,0)$, $q=4g_1$, half of the supersymmetries are
preserved \cite{SS}. In this case, the SUSY conditions for constant spinors are
\bea
\delta\chi&=&\delta(T_I\lambda^I_6)=\delta\psi_\r= \d \psi_\m=0, \\
\delta\lambda_1&=&\frac{1}{2\sqrt{2}}iq ~\s^3 \otimes(\gamma_5-1)\varepsilon=0, \\
\delta\psi_\theta&=&\bigg[\partial_\theta-i\bigg(\frac{1}{f_0}-1\bigg) \s^3 \otimes (\gamma_5-1)\bigg]\varepsilon=0.
\eea
Thus, it is obvious from the above that there exists a constant 4D
Killing Weyl spinor $\tilde{\vep}_L$, with $\tilde{\vep}=(\tilde{\vep}_L,0)^T$, which  preserves 4D ${\cal N}=1$
supersymmetry. In this case, it was shown that there exists a
chiral massless mode of gravitino \cite{SS} due to the
cancellation between spin and $U(1)_R$ connections.

On the other hand, the above SUSY transformations show that SUSY can be spontaneously broken by a nonzero flux along
the $E_6$ even without conical singularities, {\it i.e.}
$T_IF^I_{mn}\neq 0$. This is partially because there is no potential term in
the variation to cancel the gauge field related part in
\reef{la6var}. The fermion SUSY transformations for constant $\vep$ then read
\bea
\delta\chi&=&\delta\psi_\r= \d \psi_\m=0, \\
\delta\la_1 &=& -i \sqrt{2} g_1 \s^3 \otimes {\bf 1} \vep, \\
\delta(T_I\lambda^I_6)&=&T_0~ i \sqrt{2}~ g_1 \s^3 \otimes \g^5 \vep,\\
\delta \psi_\th &=& \bigg[\partial_\theta-i \left({1\over f_0}-1\right) \s^3 \otimes \g^5\bigg]   \vep.
\eea

Upon introducing conical singularities to the Salam-Sezgin solution,
by deforming the space with non-SUSY brane tensions \cite{gibbons},  it
can be seen from  \reef{chiexp}-\reef{psithexp} that the SUSY is broken completely in the bulk
due to the absence of a globally well-defined Killing spinor for the conical geometries
\cite{henne}. As will be seen in the next sections, however, the  explicit
SUSY breakdown does not mean necessarily the absence of a massless
mode of the  gravitino.

\subsection{The ``4D gravitino'' equation of motion}

For the background solution of the previous section, we can
rearrange the fermionic  part of the action \reef{faction} as \bea
e^{-1}{\cal L}_f&=& {\bar\chi}\Gamma^M{\cal D}_M\chi
+{\bar\lambda}\Gamma^M{\cal D}_M\lambda+{\bar\lambda}_6\Gamma^M{\cal D}_M\lambda_6 +{\bar\psi}_m\Gamma^{m\lambda n}{\cal D}_\lambda \psi_n +e^{-1}{\cal L}_{\rm mix}  \nonumber \\
&&+{\bar\psi}_\mu \Gamma^{\mu\nu\lambda}{\cal D}_\nu \psi_\lambda
+{\bar\psi}_\mu \Gamma^{\mu m\lambda}{\cal D}_m\psi_\lambda
\nonumber \\
&&+(-{\bar\psi}_\mu g^{\mu\lambda}\Gamma^m{\cal
D}_\lambda\psi_m+{\bar\psi}_\mu \Gamma^\mu\eta +{\rm h.c.}),
\label{mixeta} \eea where ${\cal L}_{\rm mix}$ contains the mixing
terms between spin-$\frac{1}{2}$ components. The linear
combination of the spin-$\frac{1}{2}$ fermions   \bea
\eta&=&\Gamma^\lambda\Gamma^m{\cal
D}_\lambda\psi_m+\Gamma^{mn}{\cal
D}_m\psi_n-\frac{1}{4}(\partial_m\phi)\Gamma^m\chi+i\sqrt{2}g_1
e^{-\frac{1}{4}\phi}\lambda_1
\nonumber \\
&&-\frac{1}{4\sqrt{2}}e^{\frac{1}{4}\phi}
(F_{1mn}\Gamma^{mn}\lambda_1+F^I_{6mn}\Gamma^{mn}\lambda^I_6) ,
\eea plays the role of the would-be Goldstone fermion and mixes
with the ``4D gravitino''\footnote{We use the term ``4D
gravitino'' to denote the 4D vector component of the gravitino
before dimensional reduction.} $\psi_\mu$. 
Even if supersymmetry is broken explicitly by brane sources, we
can get rid of
the mixing terms with the spin-$\frac{1}{2}$ fermions by the redefinition of the 4D
gravitino.  
Therefore, in order to get the mass spectrum of the ``4D
gravitino'', we only have to consider the second line of the
action \reef{mixeta} \be e^{-1}{\cal L}_\psi= {\bar\psi}_\mu
\Gamma^{\mu\nu\lambda}{\cal D}_\nu \psi_\lambda + {\bar\psi}_\mu
\Gamma^{\mu n \lambda}{\cal D}_n \psi_\lambda, \label{gramass} \ee
where \be {\cal D}_\nu=\partial_\nu+\frac{1}{2}\omega_{\nu\alpha
5}\Gamma^{\alpha 5} +\frac{1}{2}\omega_{\nu\alpha 6}\Gamma^{\alpha
6}, \ \ \  {\cal
D}_n=\partial_n+\frac{1}{2}\omega_{n56}\Gamma^{56}-ig_1A_{1n}, \ee
 and $\G^{\mu n \lambda}=-\Gamma^{\mu\lambda}\Gamma^n$. From this
action, we derive easily the equation of motion for the ``4D
gravitino'' as \be
\Gamma^{\mu\nu\lambda}(\partial_\nu+\frac{1}{2}\omega_{\nu\alpha
5}\Gamma^{\alpha 5} +\frac{1}{2}\omega_{\nu\alpha 6}\Gamma^{\alpha
6})\psi_\lambda
-\Gamma^{\mu\lambda}\Gamma^n(\partial_n+\frac{1}{2}\omega_{n56}\Gamma^{56}-ig_1A_{1n})\psi_\lambda=0.
\ee

Using the vielbein and the spin connection obtained in the
Appendix A, we can rewrite the gravitino equation as  \bea
0&=&W^{-1}\sigma^1\otimes\gamma^{\alpha\beta\gamma}
\d^\m_\g\Big[\partial_\beta+\frac{1}{2}W'
(\sigma^0\otimes\gamma_\beta\gamma^5\cos\theta+i\sigma^3\otimes\gamma_\beta\sin\theta)\Big]\psi_\m \nonumber \\
&&-\s^0 \otimes \gamma^{\alpha\gamma}\d^\m_\g(\sigma^1\otimes\gamma^5\cos\theta+\sigma^2\otimes{\bf 1}\sin\theta)\partial_\rho\psi_\m \nonumber \\
&&-\s^0 \otimes
\gamma^{\alpha\gamma}\d^\m_\g(-\sigma^1\otimes\gamma^5\sin\theta+\sigma^2\otimes{\bf
1}\cos\theta)\frac{1}{a}
\Big[\partial_\theta+\frac{1}{2}i\omega\sigma^0\otimes\gamma^5-ig_1A_{1\theta}\Big]\psi_\m,
\eea with \be \omega=1-a', \ \ g_1
A_{1\theta}=-n_1\bigg(\frac{1}{f_1}-1\bigg). \ee

  Using the chirality
condition $\sigma^3 \otimes {\bf 1}\psi_\m=\psi_\m$ and
$\gamma^{\alpha\beta\gamma}\gamma_\beta=-2\gamma^{\alpha\gamma}$,
we can simplify the above equation as \bea
W^{-1}\sigma^1\otimes\gamma^{\alpha\beta\gamma}\d^\m_\g\partial_\beta\psi_\m&=&\sigma^1\otimes\gamma^{\alpha\gamma}\d^\m_\g
\bigg[(\gamma^5\cos\theta+i\sin\theta)(\partial_\rho+\frac{W'}{W}) \nonumber \\
&&+(-\gamma^5\sin\theta+i\cos\theta)
\frac{1}{a}(\partial_\theta+\frac{1}{2}i\omega\gamma^5-ig_1A_{1\theta})\bigg]\psi_\m.
\eea After imposing the gauge fixing conditions,
$\Gamma^\m\psi_\m=0$ and $\partial^\m\psi_\m=0$, as well as using
the identity
$\gamma^{\alpha\beta\gamma}=\gamma^\alpha\gamma^\beta\gamma^\gamma-\eta^{\alpha\beta}\gamma^\gamma-\eta^{\beta\gamma}\gamma^\alpha
+\eta^{\alpha\gamma}\gamma^\beta$,
 the gravitino equation becomes \bea
W^{-1}\sigma^1\otimes\gamma^\beta\partial_\beta\psi_\m&=&
-\sigma^1\otimes\bigg[(\gamma^5\cos\theta+i\sin\theta)(\partial_\rho+\frac{W'}{W}) \nonumber \\
&&+(-\gamma^5\sin\theta+i\cos\theta)
\frac{1}{a}(\partial_\theta+\frac{1}{2}i\omega\gamma^5-ig_1A_{1\theta})\bigg]\psi_\m.
\eea

Finally, decomposing  the 6D Weyl spinor $\tilde{\psi}_\m$ to left
and right components as $\tilde{\psi}_\m=(\tilde{\psi}_{\m
L},\tilde{\psi}_{\m R})^T$, satisfying $\gamma^5(\tilde{\psi}_{\m
L},0)^T=+(\tilde{\psi}_{\m L},0)^T$ and $\gamma^5
(0,\tilde{\psi}_{\m R})^T=-(0,\tilde{\psi}_{\m R})^T$, we obtain
the final form of the gravitino equation as \bea
W^{-1}{\bar\sigma}^\beta\partial_\beta \tilde{\psi}_{\m
L}&=&e^{-i\theta}
\Big[\partial_\rho+\frac{W'}{W}+{1 \over a}(-i\partial_\theta-\frac{1}{2}\omega-g_1A_{1\theta})\Big]\tilde{\psi}_{\m R}, \label{grav1}\\
W^{-1}\sigma^\beta\partial_\beta \tilde{\psi}_{\m R}&=&e^{i\theta}
\Big[-\partial_\rho-\frac{W'}{W}+{1 \over
a}(-i\partial_\theta+\frac{1}{2}\omega-g_1A_{1\theta})\Big]\tilde{\psi}_{\m
L}.\label{grav2} \eea

\section{Solutions to the gravitino equation}

We will now solve the above equations of motion by dimensionally
reducing to 4D mass eigenstates. Therefore, we make a Fourier expansion of
the ``4D gravitino'' as \bea
\tilde{\psi}_{\m L}&=&\sum_m \tilde{\psi}^{(m)}_{\m L}(x)\varphi^{(m)}_L(\rho)e^{im\theta}, \\
\tilde{\psi}_{\m R}&=&\sum_m \tilde{\psi}^{(m)}_{\m
R}(x)\varphi^{(m)}_R(\rho)e^{im\theta}. \eea Then, plugging the
above Fourier expansions into eqs.(\ref{grav1}) and (\ref{grav2}),
we get the equations for the wave functions of the
gravitino\footnote{Compare to the spin-$\frac{1}{2}$ fermions
\cite{salvio} charged under $U(1)_R$, for which the warp factor
dependence in the equation comes as $2W'/W$ instead of $W'/W$.}
\bea W^{-1}{\bar\sigma}^\beta\partial_\beta
\tilde{\psi}^{(m-1)}_{\m L}\varphi^{(m-1)}_L&=&
\Big[\partial_\rho+\frac{W'}{W}+{1 \over a}(m-\frac{1}{2}\omega-g_1A_{1\theta})\Big]\tilde{\psi}^{(m)}_{\m R}\varphi^{(m)}_R, \label{grav1f}\\
W^{-1}\sigma^\beta\partial_\beta \tilde{\psi}^{(m+1)}_{\m
R}\varphi^{(m+1)}_R&=& \Big[-\partial_\rho-\frac{W'}{W}+{1 \over
a}(m+\frac{1}{2}\omega-g_1A_{1\theta})\Big] \tilde{\psi}^{(m)}_{\m
L}\varphi^{(m)}_L. \label{grav2f} \eea The KK massive modes of
gravitino are satisfying \bea
{\bar\sigma}^\beta\partial_\beta \tilde{\psi}^{(m)}_{\m L}&=&M_m \tilde{\psi}^{(m+1)}_{\m R}, \nonumber\\
\sigma^\beta\partial_\beta \tilde{\psi}^{(m+1)}_{\m R}&=&M_m
\tilde{\psi}^{(m)}_{\m L}, \label{mass4d} \eea with $M_m$ the KK
mass of each 4D Dirac gravitino
$\tilde{\psi}^{(m)}_{\m}=(\tilde{\psi}^{(m)}_{\m
L},\tilde{\psi}^{(m+1)}_{\m R})^T$. Then, the equations for the
gravitino wave functions become \bea
W^{-1}M_{m-1} \varphi^{(m-1)}_L&=&\Big[\partial_\rho+\frac{W'}{W}+{1 \over a}(m-\frac{1}{2}\omega-g_1A_{1\theta})\Big]\varphi^{(m)}_R, \label{phim1}\\
W^{-1}M_m\varphi^{(m+1)}_R&=& \Big[-\partial_\rho-\frac{W'}{W}+{1
\over a}(m+\frac{1}{2}\omega-g_1A_{1\theta})\Big]\varphi^{(m)}_L.
\label{phim2}\eea

The normalizability condition for the
gravitino is \bea \int d\theta \int d\rho ~
Wa~ | \varphi^{(m)}_{L,R}|^2<\infty. \label{nc} \eea

Furthermore, there is a hermiticity condition for the gravitino,
which, in analogy with \cite{salvio}, is given by \be
 \int d^6x ~\de_N(\sqrt{-G} ~\bar{\psi}_M \G^{MN \La} \psi_\La)=0.
\ee
In terms of the above mode decomposition, this gives
\bea
W^2a~ \varphi^{(m)*}_L\varphi^{(m+1)}_R\Big|^{\rho_s}_0=0.\label{hc}
\eea

\subsection{Massless modes}

For massless modes, we set ${\bar\sigma}^\beta\partial_\beta
\tilde{\psi}^{(m)}_{\m L}=\sigma^\beta\partial_\beta
\tilde{\psi}^{(m)}_{\m R}=0$ in \reef{grav1f}, \reef{grav2f}.
Then, the equations of left-handed and right-handed gravitinos are
decoupled as \bea
\Big[\partial_\rho+\frac{W'}{W}+{1 \over a}(m-\frac{1}{2}\omega-g_1A_{1\theta})\Big]\varphi^{(m)}_R&=&0, \\
\Big[\partial_\rho+\frac{W'}{W}+{1 \over
a}(-m-\frac{1}{2}\omega+g_1A_{1\theta})\Big]\varphi^{(m)}_L&=&0.
\eea We can find the explicit solution of the above equations as
\bea
\varphi^{(m)}_L&=& {1 \over W }  ~{\rm exp}\bigg[\int^\rho d\rho' {1 \over a }(m+\frac{1}{2}\o-g_1A_{1\theta})\bigg] \nonumber \\
&=&{N_m \over W\sqrt{a}}~\Big(\frac{r}{r_0}\Big)^{\frac{s}{2}}~f_0^{\frac{1-t}{2}}, \label{zerow}
\eea
with
\bea
s&=&\frac{1}{\lambda}(1+2m), \nonumber \\
t&=&\frac{1}{\lambda}(m+\frac{1}{2}-n_1)\Big(1-\frac{r^2_0}{r^2_1}\Big)+\frac{n_1}{\lambda}+1,
\eea
where $N_m$ is the normalization constant.  In the above, we have used that
\be
\int dr {f_1 \over r f_0} \sim  \ln [r f_0^{r_0^2-r1^2 \over 2r_1^2}],~~~~
\int dr {1 \over rf_0} \sim  \ln [r f_0^{-1/2}].
\ee
The solution for the right-handed gravitino is given by the one for the left-handed gravitino (\ref{zerow}) with $(m,n_1)$ being replaced by $(-m,-n_1)$.

From the normalization condition (\ref{nc}), we determine the
normalization constant of the general solution (\ref{zerow}) as
\bea N^2_m=\frac{1}{2\pi
r_0}\bigg(\int^\infty_0\frac{x^s}{(1+x^2)^t}\bigg)^{-1}\equiv\frac{\Gamma_m}{2\pi
r_0},
\eea with
\be
\Gamma_m\equiv\frac{2\Gamma[t]}{\Gamma[(1+s)/2]\Gamma[t-(1+s)/2]}.
\ee
In order to have  finite norm, we require the following
inequalities for the existence of left-handed zero mode
\be
s>-1, \quad s-2t<-1,
\ee
or in terms of our original
parameters \bea - \frac{1}{2}(1+\lambda) < m < n_1-\frac{1}{2}
\left( 1 - \lambda \frac{r^2_1}{r^2_0}\right)~. \label{norm} \eea
For the right-handed zero mode, the normalizability conditions are
\bea
 n_1+\frac{1}{2} \left( 1 - \lambda \frac{r^2_1}{r^2_0}\right) < m < \frac{1}{2}(1+\lambda) .
\label{norm1}
\eea

Let us first discuss about the simple cases with constant warp
factor. In the sphere case with $U(1)_R$ flux only, which is the
Salam-Sezgin solution, we have the relation,
$g_1A_{1\theta}=\frac{1}{2}\omega$. In the above general
expression, we take $r_0=r_1$, $\lambda=1$ and $n_1=1$. Then, from
the normalizability conditions (\ref{norm}) and (\ref{norm1}), we
obtain only one massless mode (for $m=0$) from the left-handed
gravitino as \bea \varphi^{(0)}_L&\propto&a^{-1/2}~ r^{{1 \over
2}}~f_0^{-1/2}={\rm constant}. \eea
On the other hand, in the
sphere case with $E_6$ flux only, {\it i.e.} $A_{1\theta}=0$, we
take $r_0=r_1$, $\lambda=1$ and $n_1=0$. Thus, from
eqs.~(\ref{norm}) and (\ref{norm1}), we can see that there is no
normalizable massless mode for any $m$. This shows that
supersymmetry is completely broken.

Now let us look at the effects of a nonzero deficit angle and/or a warp factor.
In this case, it is possible to have both $U(1)_R$ and $E_6$ fluxes non-vanishing.
In the unwarped case with $\lambda\neq 1$ and $r_0=r_1$, the wave functions of massless modes are
\bea
\varphi^{(m)}_R&\propto&   a^{-1/2}~ r^{{1 \over \lambda}\left({1 \over 2}-m\right)}~f_0^{\frac{n_1}{2\lambda}}  , \\
\varphi^{(m)}_L&\propto&  a^{-1/2}~ r^{{1 \over \lambda}\left({1
\over 2}+m\right)}~f_0^{-\frac{n_1}{2\lambda}}  . \eea Since
$\lambda=\sqrt{n^2_1+n^2_6g^2_1/g^2_6} \geq |n_1|$, for $n_1>0$,
we can see that there is at least one massless mode of
$\varphi^{(m)}_L$ from eq.~(\ref{norm}). Thus, although
supersymmetry is broken by the $E_6$ flux and also explicitly by
the brane sources, there exist massless modes of the gravitino.
For example, in the  case of $n_6=0$, $n_1=\la$, $q=4g_1$, there
exist normalizable modes for the left-handed gravitino and the
nontrivial SUSY variation \reef{psithexp} becomes \be \d
{\tilde\psi}_{\th L}=
\Big[\partial_\theta-\frac{1}{2}i(n_1-1)\Big] {\tilde\vep}_L. \ee
Thus, $\d {\tilde\psi}_{\theta L}=0$ would require
${\tilde\vep}_L\propto e^{i\frac{1}{2}(n_1-1)\theta}$. For even
$n_1$, the Killing spinor is not single-valued so there is no
remaining SUSY in this case. On the other hand, for odd $n_1$,
there is an $N=1$ SUSY left in the bulk even if SUSY is broken
explicitly by the deficit angle at the brane.

Furthermore, in the case with the non-constant warp factor, we
also find, from the $r^2_1/r^2_0$ term in eq.~(\ref{norm}),  that
the massless modes are still maintained. Therefore, as far as
$U(1)_R$ flux is nonzero, the massless mode of gravitino obtained
in the Salam-Sezgin solution remains even with a nonzero deficit
angle or a warp factor. This property of the appearance of
massless modes when $U(1)_R$ is present, holds also for  for bulk
spin-$\frac{1}{2}$ fermions which are charged under the $U(1)_R$.
Then it has been shown in \cite{RSS,salvio} that massless modes of
these fermions persist after flux compactification. For the
non-constant warp factor, however, there is no remaining SUSY in
the bulk at all as shown from the SUSY transformations in Section
3.1.

\subsection{Massive modes}

Returning to the massive modes, we can substitute
$\varphi^{(m+1)}_R$ from \reef{phim2} to \reef{phim1} and obtain a
single second order differential equation for $\varphi^{(m)}_L$
\bea M^2_m\varphi^{(m)}_L&=&
W\Big[\partial_\rho+\frac{W'}{W}+{1 \over a}(m+1-\frac{1}{2}\omega-g_1A_{1\theta})\Big]\nonumber  \\
&&\quad\times W\Big[-\partial_\rho-\frac{W'}{W}+{1 \over
a}(m+\frac{1}{2}\omega-g_1A_{1\theta})\Big]\varphi^{(m)}_L,
\label{singlediff} \eea supplemented by eq. \reef{phim2} which
acts like  a constraint equation. From the latter equation, the
hermiticity condition (\ref{hc}) becomes \be
W^3a\varphi^{(m)*}_L\bigg[-\partial_\rho-\frac{W'}{W}+(m+\frac{1}{2}\o-g_1A_{1\theta})/a\bigg]\varphi^{(m)}_L\Big|^{\rho_s}_0=0.
\ee

Before tackling the most general case, let us present the
solutions for the massive modes in the two simple cases of the
supersymmetric (SS vacuum) and the non-supersymmetric (with $E_6$
flux) solutions. In both cases the internal space is a sphere and
there is no warping.

\subsubsection{Massive modes for the Salam-Sezgin solution}

 First, for the sphere case with a $U(1)_R$ flux only,
$g_1A_{1\theta}=\frac{1}{2}\omega$ so the equation for massive
modes becomes \be
\bigg[\partial^2_\rho+\frac{a'}{a}\partial_\rho-\frac{m^2}{a^2}+M^2_m\bigg]\varphi^{(m)}_L=0,
\ee where $a=\frac{r_0}{2}\sin(\frac{2\rho}{r_0})$. Then, by
making a change of variables as $y=\cos(\frac{2\rho}{r_0})$, the
above equation can be cast into \be
(1-y^2)\frac{d^2\varphi^{(m)}_L}{dy^2}-2y\frac{d\varphi^{(m)}_L}{dy}
+\bigg(-\frac{m^2}{1-y^2}+\frac{1}{4}r^2_0M^2_m\bigg)\varphi^{(m)}_L=0.
\ee This is nothing but the Legendre's associated differential
equation. So, we can find that the KK mass spectrum  is the one
for the spherical harmonics \be M^2_{m,n}=\frac{4}{r^2_0} n(n+1),
\ \ \ n=1,2,\cdots, \ee with degeneracies $|m|\leq n$. The wave
functions of massive modes are given by \be
\varphi^{(m,n)}_L=P^m_n(y)=\frac{(1-y^2)^{m/2}}{2^n
n!}\frac{d^{m+n}}{dy^{m+n}}(y^2-1)^n. \ee

\subsubsection{Massive modes for the non-supersymmetric sphere solution}

In the sphere case with $E_6$ flux only, $A_{1\theta}=0$, so the
equation for massive modes is \be
\bigg[\partial^2_\rho+\frac{a'}{a}\partial_\rho-\frac{(m+\frac{1}{2})^2}{a^2}
-\frac{a^{\prime
2}}{4a^2}+\frac{a^{\prime\prime}}{2a}+\frac{(m+\frac{1}{2})a'}{a^2}+M^2_m\bigg]\varphi^{(m)}_L=0,
\ee where $a=\frac{r_0}{2}\sin(\frac{2\rho}{r_0})$. Then, with the
field redefinition \be
{\tilde\varphi}^{(m)}=\sqrt{a}~\varphi^{(m)}_L, \ee the above
equation becomes \be
(-\partial^2_\rho+V(\rho)){\tilde\varphi}^{(m)}=M^2_m{\tilde\varphi}^{(m)},
\ee where \be
r^2_0V=v_0+v_1\tan^2(\frac{\rho}{r_0})+v_2\cot^2(\frac{\rho}{r_0}),
\ee with \be v_0=2(m+\frac{1}{2})^2, \ \ v_1=(m+1)^2-\frac{1}{4},
\ \ v_2=m^2-\frac{1}{4}. \label{vdef} \ee Next, making a change of
variables and doing another field redefinition \cite{salvio} as
\bea
z&=&\cos^2(\frac{\rho}{r_0}), \\
{\tilde\varphi}^{(m)}&=&z^\gamma (1-z)^\beta \psi^{(m)}(z), \eea
with \be v_1=4\gamma^2-2\gamma, \ \ v_2=4\beta^2-2\beta,
\label{para} \ee we obtain the final form of the massive modes
equation as \be
z(1-z)\frac{d^2\psi^{(m)}}{dz^2}+[c-(1+a+b)z]\frac{d\psi^{(m)}}{dz}-ab\psi^{(m)}=0,
\label{hypergeo} \ee where \bea
a&=&\beta+\gamma+\frac{1}{2}r_0M_m, \\
b&=&\beta+\gamma-\frac{1}{2}r_0M_m, \\
c&=&\frac{1}{2}+2\gamma.
\eea
In order to express the parameters $a$, $b$, $c$ in terms of the the integer $m$, we should solve the quadratic equations \reef{para} with the definitions \reef{vdef}. There are two solutions for $\b$ and $\g$, but there is no physical difference between the various choices of the solutions. Thus, from now on we make the choice
\be
\beta=\frac{1}{2}\Big(m+\frac{1}{2}\Big), \ \ \gamma=-\frac{1}{2}\Big(m+\frac{1}{2}\Big). \label{bgchoice}
\ee

From the boundary conditions for the gravitino given in the Appendix B,
we can now obtain the quantized KK masses of the gravitino.
For $\gamma\geq \frac{1}{4}$ and $\beta<\frac{1}{4}$,
we need $c-a=-n$ or $c-b=-n$ with $n=0,1,2,\cdots$. Both cases yield the spectrum
\be
M^2_{m\leq -1,n}=\frac{4}{r^2_0}(n+\frac{1}{2}+\gamma-\beta)^2=\frac{4}{r^2_0}(n-m)^2, \ \ n=0,1,2,\cdots.\label{kk1}
\ee
Moreover, for $\gamma<\frac{1}{4}$ and $\beta\geq \frac{1}{4}$,
we need $1+a-c=-n$ or $1+b-c=-n$ for $n=0,1,2,\cdots$. Then, for both cases we obtain the KK spectrum as
\be
M^2_{m\geq 0,n}=\frac{4}{r^2_0}(n+\frac{1}{2}-\gamma+\beta)^2=\frac{4}{r^2_0}(n+m+1)^2, \ \ n=0,1,2,\cdots.\label{kk2}
\ee
On the other hand, the other ranges of $\gamma$ and $\beta$, as they are chosen in \reef{bgchoice}, are not possible. With the introduction of the new quantum number $n$, it is obvious that the wavefunctions should be labeled as $\tilde{\vf}^{(m,n)}$ and the 4D modes accordingly $\tilde{\psi}^{(m,n)}$.

Consequently, from eqs.(\ref{kk1}) and (\ref{kk2}), the lowest
massive modes for the left-handed gravitino are double degenerate
with KK mass, $M^2_{0,0}=M^2_{-1,0}=\frac{4}{r^2_0}=8g^2_1$. Since
one pair of left-handed and right-handed gravitinos makes up a 4D
massive Dirac gravitino from eq. (\ref{mass4d}), we find that
there are two 4D massive gravitinos at the lowest KK level,
$\tilde{\psi}_\m^{(-1,0)}=(\tilde{\psi}^{(-1,0)}_{\m
L},\tilde{\psi}^{(0,0)}_{\m R})^T$ and
$\tilde{\psi}_\m^{(0,0)}=(\tilde{\psi}^{(0,0)}_{\m
L},\tilde{\psi}^{(1,0)}_{\m R})^T$.

\subsubsection{Massive modes for the general warped solution}

In the general warped case, we can rewrite eq. \reef{singlediff} in an expanded form as
\be
(- \de_\r^2 + h \de_\r + g_m )  \vf_L^{(m)}=W^{-2}M_m^2\vf_L^{(m)},
\ee
with
\bea
h&=&-3{W' \over W}-{a' \over a}, \\
g_m&=&-{W'' \over W}-{W'^2 \over W^2}-{1 \over 2}{a'' \over a}+{1 \over 4}{a'^2 \over a^2}-{3 \over 2}{a' W' \over a W} \nonumber \\
&&+{1 \over a}\left[ \left({W' \over W}-{a' \over a}\right)\left(m +{1 \over 2}-g_1A_{1\theta}\right) - g_1A'_{1\theta} \right] \nonumber  \\
&&+{1 \over a^2}\left[ m(m+1) +{1 \over 4} + g^2_1 A^2_{1\theta} -(2m+1) g_1 A_{1\theta} \right].
\eea

Let us make now the field redefinition $\tilde{\vf}^{(m)}=W \sqrt{a}\vf_L^{(m)}$ and the radial coordinate redefinition $d \r = W d u$. The above equation can then be cast in the following Schr\"odinger form
\be
(- \de_u^2 + V )  \tilde{\vf}^{(m)}=M_m^2 \tilde{\vf}^{(m)},
\ee
with the potential given by
\bea
V&=&\partial_u\left(\frac{W}{a}\right)\left(m+{1 \over 2} \right)-g_1\de_u \left({W \over a}  A_{1\theta} \right) \nonumber \\&&+ {W^2 \over a^2}\left[ m(m+1) +{1 \over 4} + g^2_1 A^2_{1\theta} -(2m+1) g_1 A_{1\theta} \right].
\eea

Substituting the background solutions in the above equation we find that
\be
r^2_0V=v_0+v_1\tan^2(\frac{u}{r_0})+v_2\cot^2(\frac{u}{r_0}),
\ee
with
\bea
v_0&=&-\frac{1}{\lambda}\Big(m+\frac{1}{2}\Big)+\frac{1}{\lambda}\Big(m+\frac{1}{2}-n_1\Big)
\bigg[1+\frac{2}{\lambda}\Big(m+\frac{1}{2}\Big)\bigg]\frac{r^2_0}{r^2_1}, \\
v_1&=&\frac{1}{\lambda}\Big(m+\frac{1}{2}-n_1\Big)\frac{r^2_0}{r^2_1}+\frac{1}{\lambda^2}
\Big(m+\frac{1}{2}-n_1\Big)^2\frac{r^4_0}{r^4_1}, \label{v1}\\
v_2&=&-\frac{1}{\lambda}\Big(m+\frac{1}{2}\Big)+\frac{1}{\lambda^2}\Big(m+\frac{1}{2}\Big)^2.
\label{v2}
\eea
Then, making similar redefinitions as in the non-supersymmetric sphere solution,
\bea
z&=&\cos^2(\frac{u}{r_0}), \\
{\tilde\varphi}^{(m)}&=&z^\gamma (1-z)^\beta \psi^{(m)}(z),
\eea
the parameters in the hypergeometric equation (\ref{hypergeo}) are defined as
\bea
a&=&\b +\g +{1 \over 2}\sqrt{\xi_m+r_0^2 M_m^2}\\
b&=&\b +\g -{1 \over 2}\sqrt{\xi_m+r_0^2 M_m^2}\\
c&=&{1 \over 2}+2\g
\eea
with
\be
v_1=4\gamma^2-2\gamma, \ \ v_2=4\beta^2-2\beta,\label{vrel}
\ee
From eqs.~(\ref{v1}), (\ref{v2}) and (\ref{vrel}), we can solve for $\beta$ and $\gamma$. Without loss of generality, we make the following choice of roots
\be
\beta=\frac{1}{2\lambda}\Big(m+\frac{1}{2}\Big),  \ \ \
\gamma=-\frac{1}{2\lambda}\Big(m+\frac{1}{2}-n_1\Big)\frac{r^2_0}{r^2_1}.
\ee
Furthermore, unlike the non-supersymmetric sphere solution, the following quantity is non-vanishing:
\bea
\xi_m&=&v_1+v_2-v_0 \nonumber \\
&=&\frac{1}{\lambda^2}\bigg[m+\frac{1}{2}-\frac{r^2_0}{r^2_1}\Big(m+\frac{1}{2}-n_1\Big)\bigg]^2 \nonumber \\
&=&4(\beta+\gamma)^2.
\eea

Using the boundary conditions at $z=0$ and $z=1$ as described  in the Appendix B,
we find the KK spectrum depending on the parameters $\beta$ and $\gamma$:\\
For $\gamma\geq \frac{1}{4}$ and $\beta<\frac{1}{4}$,
\bea
r^2_0 M^2_{m,n}&=&4(n+\frac{1}{2}+\gamma-\beta)^2-\xi_m \nonumber \\
&=&4(n+\frac{1}{2}+2\gamma)(n+\frac{1}{2}-2\beta).
\eea
For $\gamma\geq \frac{1}{4}$ and $\beta\geq \frac{1}{4}$,
\bea
r^2_0 M^2_{m,n}&=&4(n+\beta+\gamma)^2-\xi_m \nonumber \\
&=&4n(n+2\beta+2\gamma).
\eea
For $\gamma<\frac{1}{4}$ and $\beta<\frac{1}{4}$,
\bea
r^2_0M^2_{m,n}&=&4(n+1-\beta-\gamma)^2-\xi_m \nonumber \\
&=& 4(n+1)(n+1-2\beta-2\gamma).
\eea
For $\gamma<\frac{1}{4}$ and $\beta\geq\frac{1}{4}$,
\bea
r^2_0 M^2_{m,n}&=&4(n+\frac{1}{2}-\gamma+\beta)^2-\xi_m \nonumber \\
&=& 4(n+\frac{1}{2}+2\beta)(n+\frac{1}{2}-2\gamma).
\eea
In all the above case, we take $n=0,1,2,\cdots$.
We note that even with a nonzero $\xi_m$, the KK mass squared is always positive. Finally, it is understood that with the introduction of the quantum number $n$, the wavefunctions are labeled as $\tilde{\vf}^{(m,n)}$ and the 4D modes accordingly $\tilde{\psi}^{(m,n)}$.

\section{The brane-induced gravitino mass term}

In the previous section, we have seen that even when SUSY is
broken explicitly by the presence of brane tensions, there remains
at least one  massless mode of the gravitino in the case with
$U(1)_R$ flux. One way to generate mass for this lowest gravitino
mode, is to include a brane-induced gravitino mass term.

It has been known that a brane mass term for a bulk field on a
codimension-two brane gives rise to the divergence of the
propagator of the bulk field even at tree level \cite{GW} (see
also \cite{Dudas} for a torus orbifold case). This classical
divergence has to do with the assumption of an infinitely thin
brane.  This should have been expected, since it is known that it
is not possible to accommodate normal matter (other than the brane
tension) on a codimension-two brane in Einstein gravity
\cite{ClineGiova}. Therefore, it is imperative to regularize the
brane by introducing some thickness \cite{thickb}.

In this section, we compute the mass generation for the gravitino
zero mode,  when the brane is regularized according with  the
approach suggested in \cite{tps}. In this regularization, the
conical singularity is cut out and replaced by a spherical cap and
a ring-like brane is situated at the boundary of the spherical
cap. Thus, the regularized brane has four spatial dimensions one
of which is compactified on a circle. This procedure  has been
extended to the warped solution of  6D Salam-Sezgin supergravity
in \cite{ppz}.

Let us now discuss the effect of a brane-induced gravitino mass
term. If the ring is  located  at $\rho=\delta$  (instead of the
codimension-two brane original position  $\rho=0$), then the
induced metric on the ring brane is \be
ds^2_5=W^2(\delta)\eta_{\mu\nu}dx^\mu
dx^\nu+a^2(\delta)d\theta^2\equiv
h_{\hat{\m}\hat{\n}}dx^{\hat{\m}} dx^{\hat{\n}}, \ee with
$\hat{\m},\hat{\n}=0,1,2,3,5$. The left-handed gravitino always
has a massless mode for a nonzero monopole number of the $U(1)_R$
and in general more  than one, labeled by the winding number $m$.
Let us consider
a Majorana mass term for the left-handed gravitino only
on the regularized brane as
\be
{\cal L}_{\rm brane}=- \lambda_0
~\bar{\xi}_{\hat{\m}}\gamma^{\hat{\m} \hat{\n}}\xi_{\hat{\n}} ~
\delta(\rho-\delta), \ee where $\lambda_0$ is a dimensionless
parameter and $\xi_{\hat{\m}}$ is the 5D Majorana gravitino that
is composed of the bulk left-handed gravitino
$\tilde{\psi}_{\hat{\m} L}$ as $\xi_{\hat{\m}}\equiv
(\tilde{\psi}_{\hat{\m}L},\bar{\tilde{\psi}}_{\hat{\m}L})^T$.
Since, we have considered $\tilde{\psi}_{{\hat\mu}L} =\sum_m
\vf_L^{(m)}(\r) \tilde{\psi}_{\hat{\m}L}^{(m)}e^{im\th} \equiv
\sum_m \chi^{(m)}_{{\hat\mu}L}$, we can rewrite the gravitino mass
term in terms of the 4D Weyl spinors $\chi^{(m)}_{\hat{\m} L}$ as
\be {\cal L}_{\rm brane}=-\lambda_0~[\sum_{m,m'}(\chi^{(m)}_{\mu
L}\sigma^{[\mu}{\bar\sigma}^{\nu]}\chi^{(m')}_{\nu L} +
{\bar\chi}^{(m)}_{\mu
L}{\bar\sigma}^{[\mu}\sigma^{\nu]}{\bar\chi}^{(m')}_{\nu
L})+\cdots]~\delta(\rho-\delta), \label{bmass} \ee where the
ellipsis contains the mixing between $\chi^{(m)}_{\mu L}$ and
$\chi^{(m)}_{\theta L}$ which can be absorbed by the 4D gravitino
as in the bulk Lagrangian.

Ignoring the mixing between KK modes, we focus on the mass term
for the massless mode. We plug the wave function of the massless
mode of the left-handed gravitino (\ref{zerow}) into
eq.~(\ref{bmass}) and integrate the angle on the ring brane. Then,
we can see that only the massless mode with a zero winding number
gets a nonzero mass. If in the $r$ coordinate  the location of the
ring brane corresponding to $\r=\d$ is $r=1/\Delta_0$,  we obtain
the 4D effective gravitino mass as \be {\cal L}_{\rm
eff}=-m_{3/2}{\bar\Psi}_\mu\gamma^{\mu\nu}\Psi_\nu, \ee with
$\Psi_\mu=(\tilde{\psi}^{(0)}_{\mu
L},\bar{\tilde{\psi}}^{(0)}_{\mu L})^T$ and \be
m_{3/2}=\frac{\lambda_0\Gamma_0}{r_0}\frac{(r_0\Delta_0)^{-s}}{(1+1/(r_0\Delta_0)^2)^{t-1}}.
\ee For $r_0\Delta_0\gg 1$, the gravitino mass becomes \be
m_{3/2}\approx\frac{\lambda_0\Gamma_0}{r_0}(r_0\Delta_0)^{-s}. \ee
Since $s=\frac{1}{\lambda}>0$ for $m=0$, compatible with
normalizability condition $s>-1$, the gravitino mass is suppressed
by the delocalization of the zero mode at the regularized brane at
$r=1/\Delta_0$, compared to the compactification scale $1/r_0$.

Similarly, for the gravitino mass term with $\lambda_\infty$
at the ring brane located at $r=r^2_0\Delta_\infty$,
we obtain the 4D effective gravitino mass as
\bea
m_{3/2}=\frac{\lambda_\infty\Gamma_0}{r_0} \frac{(r_0\Delta_\infty)^s}{(1+(r_0\Delta_\infty)^2)^{t-1}}.
\eea
Thus, for $r_0\Delta_\infty\gg 1$, the gravitino mass becomes
\be
m_{3/2}\approx \frac{\lambda_\infty\Gamma_0}{r_0} (r_0\Delta_\infty)^{s-2t+2}.
\ee
If we take  $s-2t+2=\frac{1}{\lambda}\frac{r^2_0}{r^2_1}(1-2n_1)<0$ or $n_1>\frac{1}{2}$, compatible with the normalizability condition $s-2t<-1$, we can also get a suppression of the gravitino mass due to the delocalization of the zero mode at the regularized brane at $r=r^2_0\Delta_\infty$.
Therefore, for the zero-mode gravitino with the zero winding number,
we find that gravitino mass terms localized at both regularized branes give rise to a suppressed effective gravitino mass.

For both of the above cases, if we consider the zero thickness
limit of $\D_{0,\infty} \to \infty$, the effective gravitino mass
vanishes. Therefore, we see that the nonzero gravitino mass
crucially depends on the regularization of the brane. In other
words, a nonzero gravitino mass depends on the mechanism to
stabilize the brane thickness.

Finally, let us comment on the possibility of giving a mass to the
massless mode of the gravitino by some bulk operator. A plausible
case would be from the quartic fermion terms provided in \cite{NS},
if gaugino condensation is realized.
For instance, since $E_7$ gauginos are charged under $U(1)_R$,
they will have massless modes at
the compactification scale which can serve for condensation and
subsequent gravitino mass generation in the 4D effective theory.
Then, one should be careful with the possible backreaction to the
background solution of the additional terms (see \cite{abpq}) in the
scalar potential that will arise after condensation, although as noted in
\cite{NS} this will be absent for constant dilaton and gravitino
wavefunctions.

\section{Conclusions}

In the present paper, we  discussed the spectrum of the gravitino
of the six-dimensional  gauged supergravity model with gauge group
$E_7 \times E_6 \times U(1)_R$, where  a gauge flux is turned on
in the $U(1) \subset E_6$ and the $U(1)_R$ directions. We studied
in detail the spectrum in the general warped background where
codimension-two branes were supporting the necessary conical
singularities.

An interesting  result of the paper is that there exist zero modes
in the gravitino spectrum even in the case where supersymmetry is
explicitly broken by the presence of the non-supersymmetric
branes. This seems to be a special property whenever magnetic flux
lies  in the $U(1)_R$ direction. In order to give mass to these
massless modes, the explicit supersymmetry breaking by the brane
tensions is not enough, and further bulk or brane operators should
be considered.

We  have considered for simplicity, a brane operator for the
gravitino, namely a brane Majorana mass. Regularizing the brane as
in \cite{tps,ppz}, we have calculated the mass that is generated
for the zero mode of the gravitino. The resulting effective
gravitino mass depends on the localization of the wavefunction of
the zero mode and when the winding number is zero, the gravitino
mass can be parametrically smaller than the compactification
scale. This suppression, however, is power-like and if the
compactification scale is of the order of the GUT breaking scale,
we would need a mechanism to explain the smallness $\la_0$ and
thus of the SUSY breaking scale.

The  above property for the massless gravitino and its mass
suppression with extra operators, should also hold for the other
fermionic states of the spectrum which we did not consider in the
present paper. In particular the gauginos which correspond to the
directions of isometry of the internal space should have the same
feature. This procedure offers an alternative way to obtain light
(in comparison with the scale of compactification) fermions in
models with extra dimensions.

\section*{Acknowledgments}

H.M.L. is supported by the DOE Contracts DOE-ER-40682-143
and DEAC02-6CH03000.

\def\theequation{A.\arabic{equation}}
\setcounter{equation}{0}
\vskip0.8cm
\noindent
{\Large \bf Appendix A: Notations and conventions}
\vskip0.4cm
\noindent

We use the metric signature $(-,+,+,+,+,+)$ for the 6D metric. The
index conventions are the following: (1) for the Einstein indices
we use  $M,N,\cdots=0,\cdots,5,6$ for the 6D indices,
$\mu,\nu,\cdots,=0,\cdots,3$ for the 4D indices and
$m,n,\cdots=5,6$ for the internal 2D indices,  (2) for the Lorentz
indices we use $A,B,\cdots=0,\cdots,5,6$ for the 6D indices,
$\alpha,\beta,\cdots=0,\cdots,3$  for the 4D indices and
$a,b,\cdots=5,6$  for the internal 2D indices.

\section*{A-1. Gamma matrices}

We take the gamma matrices in the locally flat
coordinates\cite{SS}, satisfying
$\{\Gamma_A,\Gamma_B\}=2\eta_{AB}$, to be \bea
\Gamma_\alpha&=&\sigma^1\otimes\gamma_\alpha, \ \
\Gamma_5=\sigma^1\otimes\gamma_5, \ \ \Gamma_6=\sigma^2\otimes{\bf
1}, \eea where $\g$'s are the 4D gamma matrices with
$\gamma^2_5=1$ and $\sigma$'s are the  Pauli matrices with
$[\s^i,\s^j]=2i \ep_{ijk} \s^k$, with $i,j,k=1,2,3$, \be
\s^1 = \left(\begin{array}{ll}0 & 1 \\
1 & 0 \end{array}\right), \ \ \ \s^2 = \left(\begin{array}{lr}0 & -i \\
i & 0 \end{array}\right), \ \ \ \s^3 = \left(\begin{array}{lr}1 & 0 \\
0 & -1 \end{array}\right).
\ee
The curved gamma matrices on the other hand are given in terms of the ones in the locally
flat coordinates as $\G^M= e^{~M}_A \G^A$ where $e^{~M}_A$ is the 6D vielbein. In addition, the  6D chirality operator is given by
\be
\Gamma_7=\Gamma_0\Gamma_1\cdots\Gamma_6=\sigma^3\otimes{\bf 1}.
\ee
The convention for 4D gamma matrices is that
\be
\g^\a = \left(\begin{array}{ll}0 & \s^\a \\
\bar{\s}^\a & 0 \end{array}\right), \ \  \g^5 = \left(\begin{array}{lr} {\bf 1} & 0 \\
0 & -{\bf 1} \end{array}\right), \ee with $\s^\a=({\bf 1}, \s^i)$
and $\bar{\s}^\a=(-{\bf 1}, \s^i)$. The chirality projection
operators are defined as $P_L=(1+\gamma^5)/2$ and
$P_R=(1-\gamma^5)/2$.

Finally, some useful quantities which we use in the text are the
following \be \G^{\a 5}={\bf 1} \otimes \g^\a \g^5, \ \ \ \G^{\a
6}= i \s^3 \otimes \g^\a, \ \ \ \G^{56}=i \s^3 \otimes \g^5. \ee

\section*{A-2. Spin connection}

For the general warped solution written in the Gaussian normal coordinate
\be
ds^2=W^2\eta_{\mu\nu}dx^\mu dx^\nu +d\rho^2+a^2d\theta^2,
\ee
the nonzero vielbein components are given by
\bea
e_\mu^{~\alpha}&=&W\delta^\alpha_\mu, \\
e_m^{~a}&=&\left(\begin{array}{ll}\cos\theta & -\sin\theta \\
\sin\theta & \cos\theta \end{array}\right)\left(\begin{array}{ll} 1 & 0 \\ 0 &
a \end{array}\right).
\eea
Therefore, the nonzero components of the spin connection are
\bea
\omega^\alpha\,_5&=&\cos\theta ~W'\delta^\alpha_\mu dx^\mu, \\
\omega^\alpha\,_6&=&\sin\theta ~W'\delta^\alpha_\mu dx^\mu, \\
\omega^5\,_6&=&(1-a')d\theta\equiv \omega d\theta, \eea where
prime denotes the derivative with respect to $\rho$.

\def\theequation{B.\arabic{equation}}
\setcounter{equation}{0}
\vskip0.8cm
\noindent
{\Large \bf Appendix B: Boundary conditions for the gravitino}
\vskip0.4cm
\noindent

The general solution to the hypergeometric differential equation (\ref{hypergeo})
is $\psi(z)$ with ${\tilde\varphi}=z^\gamma (1-z)^\beta \psi(z)$ is (we have suppressed the index $m$ in all wavefunctions):\\
For $c\neq 1$, \be {\tilde\varphi}=c_1\varphi_1+c_2\varphi_2, \ee
and for $c= 1$, \be
{\tilde\varphi}=c_1\varphi_1+c_2\varphi_1\int^{z(\rho)}
\frac{d\rho}{\varphi^2_1(\rho)}. \ee The $c_1,c_2$ are integration
constants and \bea
\varphi_1&=&z^\gamma (1-z)^\beta F(a,b,c,z), \\
\varphi_2&=&z^\gamma (1-z)^\beta z^{1-c} F(a+1-c,b+1-c,2-c,z).
\eea
Here we note that $F(a,b,c,z)$ is the hypergeometric function which has the properties:
\bea
F(a,b,c,z)\rightarrow 1 \ \ {\rm for} \ z\rightarrow 0,
\eea
and
\bea
F(a,b,c,z)&=&C_1 F(a,b,a+b-c+1,1-z) \nonumber \\
&&\quad+C_2(1-z)^{c-a-b}F(c-a,c-b,c-a-b+1,1-z), \eea with \be
C_1=\frac{\Gamma(c)\Gamma(c-a-b)}{\Gamma(c-a)\Gamma(c-b)}, \ \
C_2=\frac{\Gamma(c)\Gamma(-c+a+b)}{\Gamma(a)\Gamma(b)}. \ee

Now we consider the boundary conditions for the wave functions at $z=0$ and $z=1$.
First, for $c\neq 1$, as $z\rightarrow 0$, the wave function goes like
\be
{\tilde\varphi}\rightarrow c_1 z^\gamma +c_2 z^{\frac{1}{2}-\gamma}.
\ee
So, the normalizability condition gives, $c_1=0$ for $\gamma\leq -\frac{1}{4}$
while $c_2=0$ for $\gamma \geq \frac{3}{4}$. On the other hand, the hermiticity condition
gives, $c_1=0$ for $\gamma< \frac{1}{4}$ while $c_2=0$ for $\gamma> \frac{1}{4}$.
We can also show that $c_2=0$ for $c=1$ or $\gamma=\frac{1}{4}$.

Then, for $\gamma\geq \frac{1}{4}$, where $c_2=0$, the wave function at $z\rightarrow 1$ behaves as
\be
{\tilde\varphi}\rightarrow C_1(1-z)^\beta +C_2(1-z)^{\frac{1}{2}-\beta}.
\ee
Similarly, the normalizability condition gives, $C_1=0$ for $\beta\leq -\frac{1}{4}$
while $C_2=0$ for $\beta \geq \frac{3}{4}$. On the other hand, the hermiticity condition
gives, $C_1=0$ for $\beta< \frac{1}{4}$ while $C_2=0$ for $\beta\geq \frac{1}{4}$.
Therefore, for $\gamma\geq \frac{1}{4}$ and $\beta<\frac{1}{4}$,
$C_1=0$ gives $\Gamma(c-a)=\infty$ or $\Gamma(c-b)=\infty$, {\it i.e.} $c-a=-n$
or $c-b=-n$ for $n=0,1,2,\cdots$.
On the other hand, for $\gamma\geq \frac{1}{4}$ and $\beta\geq \frac{1}{4}$,
$C_2=0$ requires $\Gamma(a)=\infty$ or $\Gamma(b)=\infty$, {\it i.e.}
$a=-n$ or $b=-n$ for $n=0,1,2,\cdots$.

Finally, for $\gamma<\frac{1}{4}$, where $c_1=0$, the wave
function at $z\rightarrow 1$  goes like \be
{\tilde\varphi}\rightarrow
C'_1(1-z)^\beta+C'_2(1-z)^{\frac{1}{2}-\beta}, \ee where $C'_1,
C'_2$ are the ones obtained from $C_1,C_2$ with $a\rightarrow
a+1-c$, $b\rightarrow b+1-c$ and $c\rightarrow 2-c$. So, again the
hermiticity condition provides the strongest constraint and  for
$\beta<\frac{1}{4}$, $C'_1=0$ requires $\Gamma(1-a)=\infty$ or
$\Gamma(1-b)=\infty$, {\it i.e.} $1-a=-n$ or $1-b=-n$ for
$n=0,1,2,\cdots$. For $\beta\geq\frac{1}{4}$, $C'_2=0$ would
require $\Gamma(1+a-c)=\infty$ or $\Gamma(1+b-c)=\infty$, {\it
i.e.} $1+a-c=-n$ or $1+b-c=-n$ for $n=0,1,2,\cdots$.


\begin{thebibliography}{99}




%\cite{Antoniadis:1998ig}
\bibitem{add}
  I.~Antoniadis, N.~Arkani-Hamed, S.~Dimopoulos and G.~R.~Dvali,
  %``New dimensions at a millimeter to a Fermi and superstrings at a TeV,''
  Phys.\ Lett.\  B {\bf 436} (1998) 257
  [arXiv:hep-ph/9804398];
  %%CITATION = PHLTA,B436,257;%%
%\cite{Arkani-Hamed:1998nn}
%\bibitem{Arkani-Hamed:1998nn}
  N.~Arkani-Hamed, S.~Dimopoulos and G.~R.~Dvali,
  %``Phenomenology, astrophysics and cosmology of theories with  sub-millimeter
  %dimensions and TeV scale quantum gravity,''
  Phys.\ Rev.\  D {\bf 59} (1999) 086004
  [arXiv:hep-ph/9807344].
  %%CITATION = PHRVA,D59,086004;%%


%\cite{Randall:1999ee}
\bibitem{rs}
  L.~Randall and R.~Sundrum,
  %``A large mass hierarchy from a small extra dimension,''
  Phys.\ Rev.\ Lett.\  {\bf 83} (1999) 3370
  [arXiv:hep-ph/9905221].
  %%CITATION = PRLTA,83,3370;%%


%\cite{Weinberg:1988cp}
\bibitem{weinberg}
  S.~Weinberg,
  %``The cosmological constant problem,''
  Rev.\ Mod.\ Phys.\  {\bf 61} (1989) 1.
  %%CITATION = RMPHA,61,1;%%


\bibitem{selftune}
%\cite{Arkani-Hamed:2000eg}
%\bibitem{Arkani-Hamed:2000eg}
  N.~Arkani-Hamed, S.~Dimopoulos, N.~Kaloper and R.~Sundrum,
  %``A small cosmological constant from a large extra dimension,''
  Phys.\ Lett.\  B {\bf 480} (2000) 193
  [arXiv:hep-th/0001197];
  %%CITATION = PHLTA,B480,193;%%
%\cite{Kachru:2000hf}
%\bibitem{Kachru:2000hf}
  S.~Kachru, M.~B.~Schulz and E.~Silverstein,
  %``Self-tuning flat domain walls in 5d gravity and string theory,''
  Phys.\ Rev.\  D {\bf 62} (2000) 045021
  [arXiv:hep-th/0001206];
  %%CITATION = PHRVA,D62,045021;%%
%\cite{Kim:2000mc}
%\bibitem{Kim:2000mc}
  J.~E.~Kim, B.~Kyae and H.~M.~Lee,
  %``A model for self-tuning the cosmological constant,''
  Phys.\ Rev.\ Lett.\  {\bf 86} (2001) 4223
  [arXiv:hep-th/0011118].
  %%CITATION = PRLTA,86,4223;%%


\bibitem{deficit}
J.~W.~Chen, M.~A.~Luty and E.~Ponton,
%``A critical cosmological constant from millimeter extra dimensions,''
JHEP {\bf 0009} (2000) 012
[arXiv:hep-th/0003067].
%%CITATION = HEP-TH 0003067;%%


\bibitem{6dself}
S.~Randjbar-Daemi, A.~Salam and J.~A.~Strathdee,
  %``Spontaneous Compactification In Six-Dimensional Einstein-Maxwell Theory,''
  Nucl.\ Phys.\ B {\bf 214} (1983) 491;
  %%CITATION = NUPHA,B214,491;%%
S.~M.~Carroll and M.~M.~Guica,
%``Sidestepping the cosmological constant with football-shaped extra
%dimensions,''
arXiv:hep-th/0302067;
%%CITATION = HEP-TH 0302067;%%
I.~Navarro,
%``Codimension two compactifications and the cosmological constant  problem,''
JCAP {\bf 0309} (2003) 004
[arXiv:hep-th/0302129].


\bibitem{finetune}
I.~Navarro,
%``Spheres, deficit angles and the cosmological constant,''
Class.\ Quant.\ Grav.\  {\bf 20} (2003) 3603
[arXiv:hep-th/0305014];
%%CITATION = HEP-TH 0305014;%%
H.~P.~Nilles, A.~Papazoglou and G.~Tasinato,
%``Selftuning and its footprints,''
Nucl.\ Phys.\ B {\bf 677} (2004) 405
[arXiv:hep-th/0309042];
%%CITATION = HEP-TH 0309042;%%
H.~M.~Lee,
%``A comment on the self-tuning of cosmological constant with deficit  angle on
%a sphere,''
Phys.\ Lett.\ B {\bf 587} (2004) 117
[arXiv:hep-th/0309050];
%%CITATION = HEP-TH 0309050;%%
%\bibitem{porrati}
J.~Garriga and M.~Porrati,
%``Football shaped extra dimensions and the absence of self-tuning,''
JHEP {\bf 0408} (2004) 028
[arXiv:hep-th/0406158].
%%CITATION = HEP-TH 0406158;%%



\bibitem{SS}
  H.~Nishino and E.~Sezgin,
  %``Matter And Gauge Couplings Of N=2 Supergravity In Six-Dimensions,''
  Phys.\ Lett.\  B {\bf 144} (1984) 187;
  %%CITATION = PHLTA,B144,187;%%
  A.~Salam and E.~Sezgin,
  %``Chiral Compactification On Minkowski X S**2 Of N=2 Einstein-Maxwell
  %Supergravity In Six-Dimensions,''
  Phys.\ Lett.\ B {\bf 147} (1984) 47.
  %%CITATION = PHLTA,B147,47;%%

 \bibitem{burgess1}
Y.~Aghababaie, C.~P.~Burgess, S.~L.~Parameswaran and F.~Quevedo,
  %``Towards a naturally small cosmological constant from branes in 6D
  %supergravity,''
  Nucl.\ Phys.\ B {\bf 680} (2004) 389
  [arXiv:hep-th/0304256].
  %%CITATION = HEP-TH 0304256;%%


\bibitem{gibbons}
G.~W.~Gibbons, R.~Guven and C.~N.~Pope,
  %``3-branes and uniqueness of the Salam-Sezgin vacuum,''
  Phys.\ Lett.\ B {\bf 595} (2004) 498
  [arXiv:hep-th/0307238];
  %%CITATION = HEP-TH 0307238;%%
%\bibitem{burgess03}
Y.~Aghababaie {\it et al.},
  %``Warped brane worlds in six dimensional supergravity,''
  JHEP {\bf 0309} (2003) 037
  [arXiv:hep-th/0308064];
  %%CITATION = HEP-TH 0308064;%%
%\bibitem{Burgess}
  C.~P.~Burgess, F.~Quevedo, G.~Tasinato and I.~Zavala,
  %``General axisymmetric solutions and self-tuning in 6D chiral gauged
  %supergravity,''
  JHEP {\bf 0411} (2004) 069
  [arXiv:hep-th/0408109].
  %%CITATION = HEP-TH 0408109;%%





\bibitem{lee}
H.~M.~Lee and C.~Ludeling,
  %``The general warped solution with conical branes in six-dimensional
  %supergravity,''
  JHEP {\bf 0601} (2006) 062
  [arXiv:hep-th/0510026].
  %%CITATION = HEP-TH 0510026;%%


\bibitem{singularsol}
A.~J.~Tolley, C.~P.~Burgess, D.~Hoover and Y.~Aghababaie,
%``Bulk singularities and the effective cosmological constant for higher
%co-dimension branes,''
arXiv:hep-th/0512218.
%%CITATION = HEP-TH 0512218;%%


\bibitem{scalarpert}
  H.~M.~Lee and A.~Papazoglou,
  %``Scalar mode analysis of the warped Salam-Sezgin model,''
  Nucl.\ Phys.\ B {\bf 747} (2006) 294
  [arXiv:hep-th/0602208],
  %%CITATION = HEP-TH 0602208;%%
  Erratum-ibid. B {\bf 765} (2007) 200;
  C.~P.~Burgess, C.~de Rham, D.~Hoover, D.~Mason and A.~J.~Tolley,
  %``Kicking the rugby ball: Perturbations of 6D gauged chiral supergravity,''
  arXiv:hep-th/0610078.
  %%CITATION = HEP-TH 0610078;%%


\bibitem{salvio}
%\cite{Parameswaran:2006db}
%\bibitem{Parameswaran:2006db}
  S.~L.~Parameswaran, S.~Randjbar-Daemi and A.~Salvio,
  %``Gauge fields, fermions and mass gaps in 6D brane worlds,''
  arXiv:hep-th/0608074.
  %%CITATION = HEP-TH/0608074;%%



\bibitem{KKLT}
  S.~Kachru, R.~Kallosh, A.~Linde and S.~P.~Trivedi,
  %``De Sitter vacua in string theory,''
  Phys.\ Rev.\  D {\bf 68} (2003) 046005
  [arXiv:hep-th/0301240].
  %%CITATION = PHRVA,D68,046005;%%



%\cite{Burgess:2005wu}
\bibitem{bulkcorr}
  C.~P.~Burgess,
  %``Supersymmetric large extra dimensions and the cosmological constant
  %problem,''
  arXiv:hep-th/0510123.
  %%CITATION = HEP-TH/0510123;%%

%\cite{Elizalde:2007di}
\bibitem{mina}
  E.~Elizalde, M.~Minamitsuji and W.~Naylor,
  %``Casimir effect in rugby-ball type flux compactifications,''
  Phys.\ Rev.\  D {\bf 75} (2007) 064032
  [arXiv:hep-th/0702098];
  %%CITATION = PHRVA,D75,064032;%%
%\cite{Minamitsuji:2007iz}
  M.~Minamitsuji,
  %``Casimir effect in a 6D warped flux compactification model,''
  arXiv:0704.3623 [gr-qc].
  %%CITATION = ARXIV:0704.3623;%%






\bibitem{RSS}
  S.~Randjbar-Daemi, A.~Salam, E.~Sezgin and J.~A.~Strathdee,
  %``An Anomaly Free Model In Six-Dimensions,''
  Phys.\ Lett.\ B {\bf 151} (1985) 351.
  %%CITATION = PHLTA,B151,351;%%






\bibitem{moresugra}
  S.~D.~Avramis, A.~Kehagias and S.~Randjbar-Daemi,
  %``A new anomaly-free gauged supergravity in six dimensions,''
  JHEP {\bf 0505} (2005) 057
  [arXiv:hep-th/0504033];
  %%CITATION = HEP-TH 0504033;%%
  S.~D.~Avramis and A.~Kehagias,
  %``A systematic search for anomaly-free supergravities in six dimensions,''
  JHEP {\bf 0510} (2005) 052
  [arXiv:hep-th/0508172];
  %%CITATION = HEP-TH 0508172;%%
  R.~Suzuki and Y.~Tachikawa,
  %``More anomaly-free models of six-dimensional gauged supergravity,''
  arXiv:hep-th/0512019.
  %%CITATION = HEP-TH 0512019;%%




\bibitem{ClineGiova}
  J.~M.~Cline, J.~Descheneau, M.~Giovannini and J.~Vinet,
  %``Cosmology of codimension-two braneworlds,''
  JHEP {\bf 0306} (2003) 048
  [arXiv:hep-th/0304147].
  %%CITATION = HEP-TH 0304147;%%




\bibitem{thickb}
M.~Kolanovic, M.~Porrati and J.~W.~Rombouts,
  %``Regularization of brane induced gravity,''
  Phys.\ Rev.\ D {\bf 68} (2003) 064018
  [arXiv:hep-th/0304148];
  %%CITATION = HEP-TH 0304148;%%
S.~Kanno and J.~Soda,
  %``Quasi-thick codimension 2 braneworld,''
  JCAP {\bf 0407} (2004) 002
  [arXiv:hep-th/0404207];
  %%CITATION = HEP-TH 0404207;%%
J.~Vinet and J.~M.~Cline,
  %``Can codimension-two branes solve the cosmological constant problem?,''
  Phys.\ Rev.\ D {\bf 70} (2004) 083514
  [arXiv:hep-th/0406141];
  %%CITATION = HEP-TH 0406141;%%
J.~Vinet and J.~M.~Cline,
  %``Codimension-two branes in six-dimensional supergravity and the
  %cosmological constant problem,''
  Phys.\ Rev.\ D {\bf 71} (2005) 064011
  [arXiv:hep-th/0501098];
  %%CITATION = HEP-TH 0501098;%%
I.~Navarro and J.~Santiago,
  %``Gravity on codimension 2 brane worlds,''
  JHEP {\bf 0502} (2005) 007
  [arXiv:hep-th/0411250];
  %%CITATION = HEP-TH 0411250;%%
C.~de Rham and A.~J.~Tolley,
  %``Gravitational waves in a codimension two braneworld,''
  JCAP {\bf 0602} (2006) 003
  [arXiv:hep-th/0511138].
  %%CITATION = HEP-TH 0511138;%%





%\cite{Peloso:2006cq}
\bibitem{tps}
  M.~Peloso, L.~Sorbo and G.~Tasinato,
  %``Standard 4d gravity on a brane in six dimensional flux compactifications,''
  Phys.\ Rev.\  D {\bf 73} (2006) 104025
  [arXiv:hep-th/0603026].
  %%CITATION = PHRVA,D73,104025;%%


%\cite{Papantonopoulos:2006dv}
\bibitem{ppz}
  E.~Papantonopoulos, A.~Papazoglou and V.~Zamarias,
  %``Regularization of conical singularities in warped six-dimensional
  %compactifications,''
  JHEP {\bf 0703} (2007) 002
  [arXiv:hep-th/0611311].
  %%CITATION = JHEPA,0703,002;%%



%\cite{Henneaux:1984ei}
\bibitem{henne}
  M.~Henneaux,
  %``Energy Momentum, Angular Momentum, And Supercharge In 2 + 1 Supergravity,''
  Phys.\ Rev.\  D {\bf 29} (1984) 2766.
  %%CITATION = PHRVA,D29,2766;%%



\bibitem{GW}
%\cite{Goldberger:2001tn}
%\bibitem{Goldberger:2001tn}
  W.~D.~Goldberger and M.~B.~Wise,
  %``Renormalization group flows for brane couplings,''
  Phys.\ Rev.\  D {\bf 65} (2002) 025011
  [arXiv:hep-th/0104170].
  %%CITATION = PHRVA,D65,025011;%%



\bibitem{Dudas}
  E.~Dudas, C.~Grojean and S.~K.~Vempati,
  %``Classical running of neutrino masses from six dimensions,''
  arXiv:hep-ph/0511001.
  %%CITATION = HEP-PH/0511001;%%


\bibitem{NS}
  H.~Nishino and E.~Sezgin,
  %``The Complete N=2, D = 6 Supergravity With Matter And Yang-Mills
  %Couplings,''
  Nucl.\ Phys.\  B {\bf 278} (1986) 353.
  %%CITATION = NUPHA,B278,353;%%



\bibitem{abpq}
  Y.~Aghababaie, C.~P.~Burgess, S.~L.~Parameswaran and F.~Quevedo,
  %``SUSY breaking and moduli stabilization from fluxes in gauged 6D
  %supergravity,''
  JHEP {\bf 0303} (2003) 032
  [arXiv:hep-th/0212091].
  %%CITATION = JHEPA,0303,032;%%






\end{thebibliography}
\end{document}